\documentclass[english,12pt]{article}
\usepackage[T1]{fontenc}
\usepackage[utf8]{inputenc}
\usepackage{geometry}
\usepackage{color}
\usepackage{babel}
\usepackage{amsmath}
\usepackage[shortlabels]{enumitem}
\usepackage{amsthm}
\usepackage{amssymb}
\usepackage{cancel}
\usepackage{esint}
\usepackage{comment}
\usepackage{hyperref}
\hypersetup{
	colorlinks,
	citecolor=blue,
	filecolor=blue,
	linkcolor=blue,
	urlcolor=blue,
	linktocpage=true
}
\usepackage[width=0.95\textwidth]{caption}
\usepackage{authblk}

\newcommand\mb[1]{\mathbb{#1}}
\newcommand\mc[1]{\mathcal{#1}}
\newcommand{\tr}{\text{Tr}}
\newcommand{\beq}{\begin{equation}}
	\newcommand{\eeq}{\end{equation}}
\definecolor{darkgreen}{rgb}{0,0.5,0}
\usepackage[rgb,dvipsnames]{xcolor}

\usepackage[titletoc]{appendix}

\usepackage{graphicx}

\usepackage{bbm}
\newcommand\id{\ensuremath{\mathbbm{1}}} 

\usepackage{bbm}
\newcommand{\be}{\begin{eqnarray}}
\newcommand{\ee}{\end{eqnarray}}
\newcommand{\bea}{\begin{eqnarray}}
\newcommand{\eea}{\end{eqnarray}}

\newcommand{\bn}{\begin{enumerate}}
\newcommand{\en}{\end{enumerate}}

\def\det{{\rm det}}



\begin{document}

\title{Global Aspects of Spaces of Vacua}
\author{Adar Sharon\footnote{adar.sharon@weizmann.ac.il}}
\affil{Department of Particle Physics and Astrophysics, Weizmann Institute of Science, Rehovot 7610001, Israel}

\maketitle

\abstract{We study "vacuum crossing", which occurs when the vacua of a theory are exchanged as we vary some periodic parameter $\theta$ in a closed loop. We show that vacuum crossing is a useful non-perturbative tool to study strongly-coupled quantum field theories, since finding vacuum crossing in a weakly-coupled regime of the theory can lead to nontrivial consequences in the strongly-coupled regime. We start by discussing a mechanism where vacuum crossing occurs due to an anomaly, and then discuss some applications of vacuum crossing in general. In particular, we argue that vacuum crossing can be used to check IR dualities and to look for emergent IR symmetries. }

\newpage
\setcounter{tocdepth}{2}
\tableofcontents{}

\newpage

\section{Introduction}

Computations in strongly-coupled quantum field theories (QFTs) are usually very challenging, since these theories cannot be studied using perturbation theory in their couplings. As a result, one is left with a very limited set of non-perturbative tools which can be used to analyze the theory. It is thus important to both expand this non-perturbative toolbox and to understand existing non-perturbative tools as much as possible.

One of these tools is the 't Hooft anomaly, which has seen a resurgence in interest in recent years. In this note we will be particularly interested in a certain form of 't Hooft anomaly, called an anomaly in the space of couplings \cite{Cordova:2019jnf,Cordova:2019uob} (see also "global inconsistencies" \cite{Kikuchi:2017pcp,Tanizaki:2018xto,Karasik:2019bxn}). A global anomaly in the space of couplings (AISC) occurs when we have some periodic parameter $\theta\sim \theta+2\pi$ which is no longer periodic when we couple to background gauge fields for some symmetry. Specifically, we expect the partition function $Z$ to obey
\begin{equation}\label{eq:AISC}\frac{Z[\theta+2\pi,A]}{Z[\theta,A]}=\exp\left(2\pi i\int \omega\right) \;,
\end{equation}
where $A$ is a background gauge field for some symmetry and the phase $\omega$ is a local functional of $A$ which cannot be removed by adding local counterterms of the gauge fields to the theory. The most interesting cases occur when $\omega$ cannot be continuously deformed to a trivial phase; this usually happens if the coefficient of $\omega$ must be quantized, or if the anomaly can be canceled by adding a $d+1$-dimensional bulk action whose coefficient must be quantized. Then this anomaly can be calculated in the UV (where the theory is usually weakly coupled), and must be matched in the IR (where it is usually strongly coupled).

\begin{figure}
	\centering
	\includegraphics[width=0.65\linewidth]{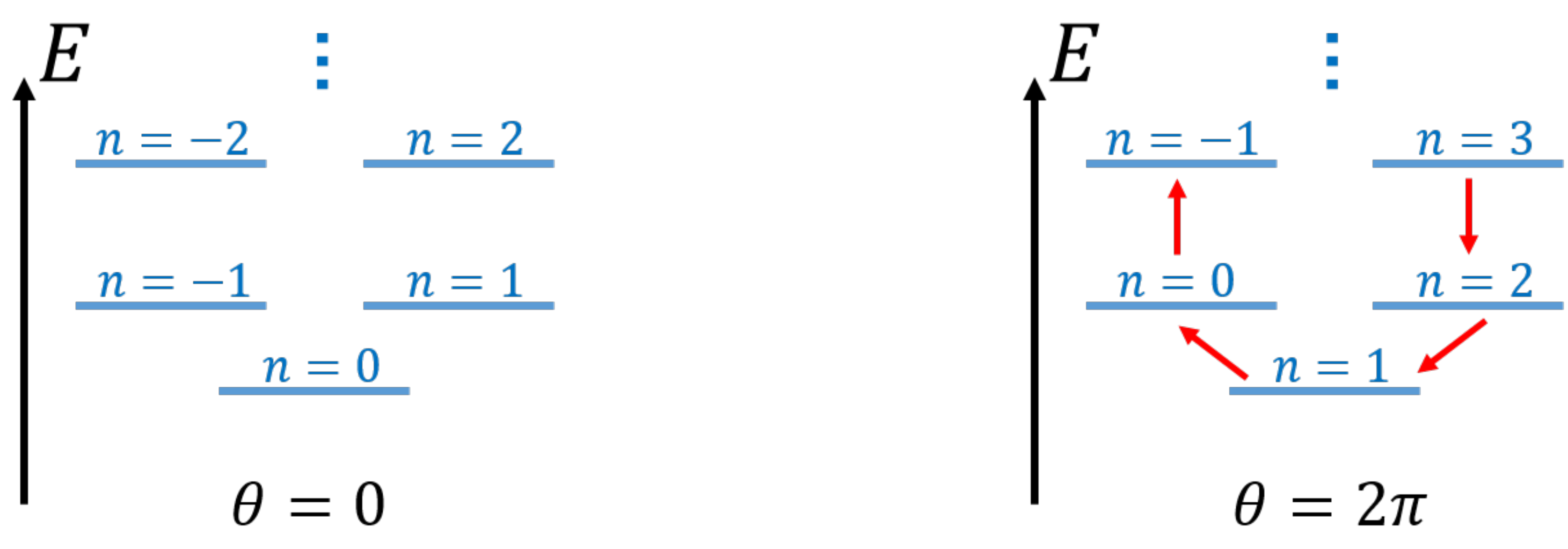}
	\caption{The energy levels for the particle on a ring are $E_n=\frac{1}{2}\left(n-\frac{\theta}{2\pi}\right)^2$ for $n\in\mathbb{Z}$. Starting at $\theta=0$ and raising it to $\theta=2\pi$, we find the same structure of energy states, but a nontrivial identification of them has occured: the $n$'th level has become the $n-1$'th level. In particular, we must have level crossing for some value of $\theta$.}
	\label{fig:particleonring}
\end{figure}

It is well known that anomalies are related to spectral flows, and in particular to level crossing.
To better understand this relation for AISCs, we restrict to the example of a quantum-mechanical\footnote{Level crossing is very different in quantum mechanics (where it usually does not occur) and in QFT. While the examples we will be interested in are QFTs, the argument is most easily understood in a quantum mechanics example, and it is easily generalized to higher dimensions.} particle on a ring (but the generalization to other AISCs is immediate), and follow an argument from \cite{Cordova:2019jnf}. Consider the theory of a particle on a circle coupled to a background gauge field $A$ for the $U(1)$ shift symmetry. The Euclidean action for this theory is
\begin{equation}
	S=\int dt \left[\frac{1}{2}( \dot q-A_t)^2-\frac{i\theta}{2\pi}(\dot q-A_t)\right]
\end{equation}
where $q\sim q+2\pi$ is compact and as a result the coupling $\theta$ is periodic, $\theta\sim \theta+2\pi$. This theory was shown to have a mixed anomaly between the $\theta$ parameter and the $U(1)$ shift symmetry \cite{Gaiotto:2017yup,Cordova:2019jnf}:
\begin{equation}
\label{eq:anomaly_particle_on_circle} \frac{Z[\theta+2\pi,A]}{Z[\theta,A]}=\exp\left(-i\int dt A_t \right) \;,
\end{equation}
which is precisely of the form \eqref{eq:AISC}. Here we recognize the $1d$ CS term on the RHS, which has a quantized coefficient. Although for a generic value of $\theta$ we expect a single ground state, we can now argue that at some value $\theta_*\in[0,2\pi]$ we must have ground state degeneracy. Indeed, since the coefficient of the CS term is quantized, a unique ground state cannot reproduce the jump \eqref{eq:anomaly_particle_on_circle}. We must therefore have level crossing at some value $\theta=\theta_*$. The energy levels for this theory can be found exactly as a function of $\theta$ at $A_t=0$, and one can explicitly show that level crossing occurs, see Figure \ref{fig:particleonring}.
This argument easily generalizes to a large class of theories and to higher dimensions. We thus expect theories with an AISC and with a gapped, trivial vacuum for generic values of a periodic parameter $\theta$ to undergo level crossing (between the vacuum and some other state) for some $\theta_*$.

\begin{figure}
	\centering
	\includegraphics[width=0.7\linewidth]{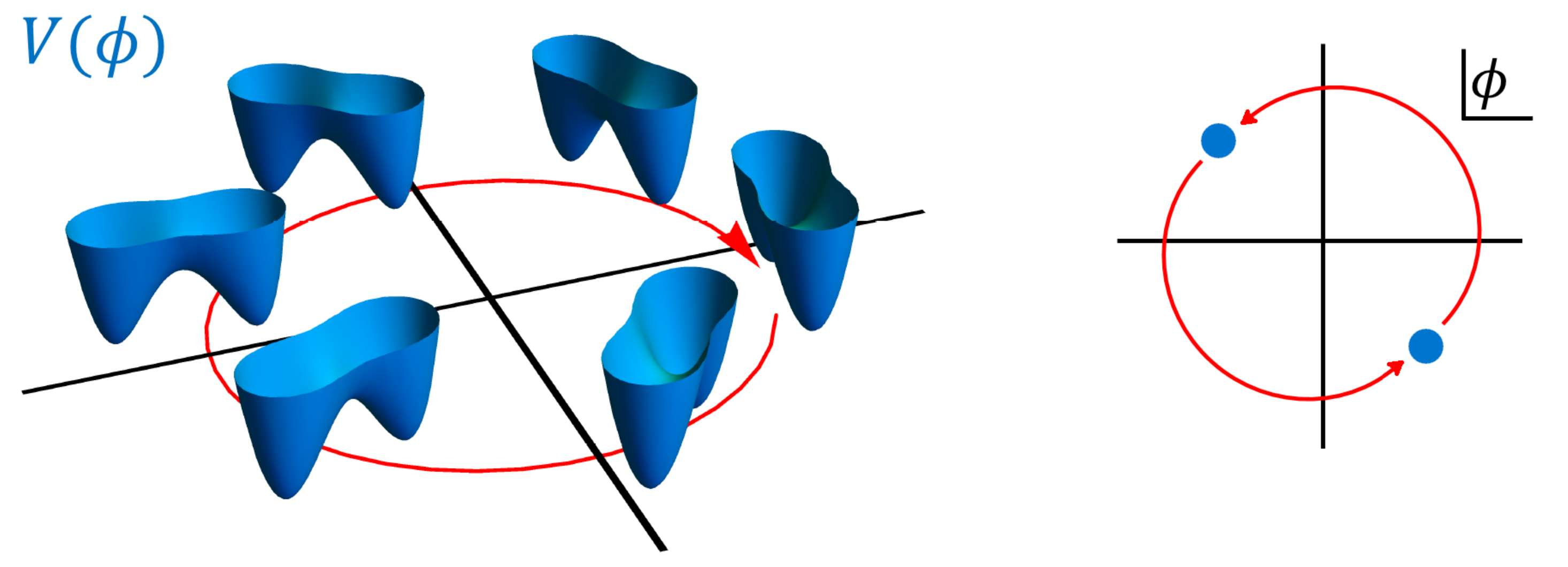}
	\caption{A schematic example of vacuum crossing. On the left we plot a potential $V(\phi)$ at various points in parameter space. The key point to note is that if we follow one minimum around a circle in parameter space, we end up in the second minimum. On the right we plot the corresponding behavior of the vacua in the $\phi$ plane. The vacua are thus exchanged once we complete the loop in parameter space.}
	\label{fig:plotpotential}
\end{figure}

It is now interesting to ask what happens if we have an AISC, but level crossing is prohibited (this can be arranged using supersymmetry, as we shall soon see). From the argument above, it is clear that this can only happen if the ground state of the theory is nontrivial, and so we will focus on theories where the vacuum is comprised of several trivial and gapped ground states which are related by some spontaneously broken discrete symmetry. Since we have prohibited level crossing, and since the argument above tells us that each ground state cannot match the anomaly on its own, it is clear that the only way to match the anomaly is by having the vacua themselves be rearranged as we take $\theta\to\theta+2\pi$. Thus, as we go from $\theta$ to $\theta+2\pi$, we find that the vacua are exchanged nontrivially. We will call this phenomenon \textbf{vacuum crossing}; it is depicted schematically in Figure \ref{fig:plotpotential}.
We emphasize that as opposed to level crossing, the states which are being exchanged here will always have the same energy.

One way to prohibit level crossing is to use supersymmetry (SUSY). Consider some SUSY theory with a nonzero Witten index and with an AISC. In addition, assume that we have a discrete number of ground states which are related by a broken (discrete) symmetry. We can now argue that level crossing with the vacua is prohibited as we vary a symmetry-preserving parameter: since the vacua are all related by the broken symmetry, they can only be "lifted" simultaneously, but the nonzero Witten index prohibits these vacua from being lifted all together (since the number of bosonic and fermionic states must be equal for all positive energies). As a result, the vacua must undergo vacuum crossing in order to match the AISC. In this note we will study some examples where vacuum crossing occurs due to this mechanism.

We can also interpret vacuum crossing as being a limiting case of level crossing, where we take the energy difference of the two levels to zero. We will discuss this interpretation in some of our examples.

It is best to have a specific (and simple) example of vacuum crossing in mind before moving on. To this end, consider a $3d$ $\mc{N}=2$ Wess-Zumino (WZ) model with a single chiral superfield $\Phi$ and superpotential $W=\frac{1}{3}\Phi^3$. This model is well understood, and in particular in the IR it flows to a conformal field theory (CFT) known as the $3d$ $\mc{N}=2$ Ising model. Next, deform the theory by adding a linear superpotential term, so that
\begin{equation} W=\frac{1}{3}\Phi^3+\lambda\Phi \;.
\end{equation}
In particular, we have explicitly broken the $U(1)_R$ R-symmetry of the model into a $\mathbb{Z}_2^R$ symmetry which takes $\Phi\to -\Phi$. We now study the vacuum structure of this new theory. For any $\lambda\neq 0$, the theory has two vacua at
\begin{equation}\label{eq:vacua_of_WZ_model}
\langle \Phi\rangle=\pm \sqrt{\lambda}\;,
\end{equation}
and so the $\mathbb{Z}_2^R$ symmetry is spontaneously broken (in some of the examples we will consider, this type of semiclassical analysis can only be trusted at large $|\lambda|$, where $\langle \Phi\rangle$ is large; otherwise, the theory is strongly coupled). The phase diagram at large $\lambda$ thus consists of two gapped vacua. We now imagine slowly going around a circle in the $\lambda$ plane, and coming back to the same point (that is, taking $\lambda\to e^{2\pi i}\lambda$). Since we start and end at the same value of $\lambda$, the overall vacuum structure must be the same. However, the two vacua do not go back to themselves - instead, they are exchanged once we complete the circle. This is easily seen by plugging in $\lambda\to e^{2\pi i}\lambda$ into the vacua \eqref{eq:vacua_of_WZ_model}, which takes $\Phi\to -\Phi$. We have thus found that although the vacuum structure has gone back to itself once we complete a circle in coupling space, there is a nontrivial identification of the vacua between themselves (see Figure \ref{fig:plotpotential}). This is vacuum crossing.

While we have argued that vacuum crossing is expected to occur in some theories with anomalies, the opposite is not necessarily true, and vacuum crossing can occur even without an AISC. It is then interesting to consider the consequences of vacuum crossing independently of AISCs. Specifically, some consequences are:
\begin{itemize}
	\item First, vacuum crossing cannot be continuously extended throughout an entire $2d$ phase diagram. We thus learn that there must be a phase transition (a singularity) for some value of $\lambda$ inside our loop. Indeed, it is known that the example discussed above has a CFT at $\lambda=0$. Since the singularity involves merging of states with the same energy, we expect this to be a continuous phase transition, and so we would always expect the singularity to be a CFT (which may be a free theory).\footnote{Another option is that the vacua escape to infinity. We will discuss this option later on.}
	\item Second, if vacuum crossing occurs close enough to a CFT, then any dual theory must also exhibit vacuum crossing. Thus vacuum crossing can be used as a check for various IR dualities. We will discuss examples where vacuum crossing is matched across a duality, and also discuss some subtelties with this rule.
	\item Finally, we will argue that vacuum crossing can be used as a tool for finding emergent IR symmetries. 
\end{itemize}
We emphasize that these consequences apply to the strongly-coupled CFT in the IR, even though the entire analysis is semiclassical and can be done solely in the weakly-coupled regime with large $|\lambda|$. We are thus able to study aspects of the strongly-coupled theory using a weakly-coupled description, and so vacuum crossing is a new non-perturbative tool to study QFTs.

Some forms of vacuum crossing have been discussed in the past in the literature, \textit{e.g.} in mass-deformed $4d$ $\mc{N}=4$ SYM (see \cite{Dorey:1999sj} and references therein). Here we will perform a more detailed study of this phenomenon and its consequences.

The rest of this note will consist of two main parts. In Section \ref{sec:examples} we will present some simple examples of vacuum crossing, and we will use them to discuss the relation between vacuum crossing and AISCs. In Section \ref{sec:applications} we will discuss vacuum crossing independently of AISCs, and study the consequences discussed above. In particular, we will argue that vacuum crossing can be used to check dualities and to look for emergent symmetries. We will give several examples to elucidate all of these points.

\section{Examples and Relation to Anomalies in Couplings}\label{sec:examples}

\subsection{$2d$ $\mathcal{N}=\left(2,2\right)$ $U(1)$ Gauge Theory}\label{sec:2d_U1_gauge_theory}

We first consider two versions of a $2d$ $\mc{N}=(2,2)$ $U(1)$ gauge theory. The first will have $N_f$ matter superfields with charge $+1$, while the second will have a single superfield of charge $q>1$. Both of these examples will exhibit vacuum crossing, which we will relate to AISCs.

First, consider a $U(1)$ gauge theory with $N_f$ matter fields of charge $+1$. We consider this theory as a function of the complex coupling $t=ir+\frac{\theta}{2\pi}$, where $r$ is the Fayet-Iliopoulos (FI) parameter\footnote{The FI parameter is renormalized in these theories, and the effective FI parameter at an energy scale $\mu$ is $r(\mu)=\frac{\sum_i q_i}{2\pi}\log(\mu/\Lambda)$ with $\{q_i\}_{i=1}^{N_f}$ the charges of the matter fields and $\Lambda$ a dynamical scale. One can imagine fixing a scale $\mu$ for the following.} and $\theta$ is the $\theta$-term. Following \textit{e.g.} \cite{Hanany:1997vm}, we find $N_f$ SUSY vacua at
\begin{equation}\label{eq:sigma_sols}
s =\mu e^{\frac{2\pi i (t+k)}{N_f}}\;,\qquad\qquad k=1,...,N_f \;,
\end{equation}
where the scalar $s$ is the bottom component of the gauge multiplet. These vacua are related to the spontaneous breaking of a discrete axial R-symmetry.\footnote{Naively these theories have a $U(1)$ axial R-symmetry, but this is broken to $\mathbb{Z}_{2N_f}$ due to an ABJ anomaly. This remaining R-symmetry is then spontaneously broken to $\mathbb{Z}_2$, leading to the $N_f$ vacua in equation \eqref{eq:sigma_sols}.} As we take $\theta$ in a loop to $\theta + 2\pi$ (which is equivalent to $t\to t+1$), we find that the vacua \eqref{eq:sigma_sols} are exchanged cyclically, and so we have vacuum crossing. 

Vacuum crossing in this theory is related to the existence of an AISC between the $\theta$ parameter and the $PSU(N_f)$ symmetry \cite{Benini:2017dus,Gaiotto:2017yup}. Explicitly, if we couple the theory to a background $PSU(N_f)$ gauge field $B$ we find 
\begin{equation} \frac{Z[\theta+2\pi,B]}{Z[\theta,B]}=\exp\left(i\frac{2\pi}{N_f}\int u_2(B) \right) \;,
\end{equation}
where $u_2(B)$ is the second Stiefel-Whitney class of the $PSU(N_f)$ bundle. Since the Witten index is $I=N_f$ (as this is just the $\mathbb{CP}^{N_f-1}$ model) and so is nonzero, we find that vacuum crossing is necessary in this theory for consistency with the AISC. 

Next, consider a $U(1)$ gauge theory with a single matter field of charge $q>1$. A similar analysis reveals $q$ SUSY vacua at
\begin{equation}\label{eq:sigma_sols_2}
s =\mu e^{\frac{2\pi i (t+k)}{q}}\;,\qquad\qquad k=1,...,q\;,
\end{equation}
which are again related to the breaking of a discrete R-symmetry.
It is easy to see that we have vacuum crossing as we take $\theta\to\theta+2\pi$. Once again, this is related to an AISC which exists in this theory \cite{Komargodski:2017dmc} (see also \cite{Cordova:2019jnf}) between the $\theta$-parameter and the $\mathbb{Z}_q$ 1-form symmetry. Indeed, coupling the theory to a background $\mathbb{Z}_q$ 2-form gauge field $K$ one finds
\begin{equation}
\label{eq:2d_U(1)_charge_q_anomaly} \frac{Z[\theta+2\pi,K]}{Z[\theta,K]}=\exp\left(-i\frac{2\pi}{q}\int K \right) \;.
\end{equation}
Since the Witten index of this theory is nonzero, we again find that vacuum crossing was necessary for consistency with the AISC.

\subsection{Wess-Zumino Models with Four Supercharges}\label{sec:WZ_model_w_4_supercharges}

Next, we consider the example from the Introduction, which is a $3d$ $\mathcal{N}=2$ Wess-Zumino (WZ) model with superpotential $W=\frac13\Phi^3+\lambda\Phi$. As discussed above, this example exhibits vacuum crossing. Similar examples can be found in other dimensions; for example, an identical analysis can be performed for a $2d$ $\mc{N}=(2,2)$ WZ model with superpotential
\begin{equation}\label{eq:simplest_superpotential_2d} W=\frac{1}{n}\Phi^n+\frac{1}{k}\lambda\Phi^k \;,
\end{equation}
for $n\geq 3$ and $k\leq n-2$. We will focus on $k=1$ for simplicity, but the result is easily generalized. At $\lambda=0$ we have a $U(1)_R$ R-symmetry, but this is explicitly broken to $\mathbb{Z}_{n-1}^R$ for any $\lambda\neq 0$. Solving the F-term equations for large $|\lambda|$ we find $n-1$ vacua due to spontaneous breaking of this $\mathbb{Z}_{n-1}^R$ symmetry. As we go around the origin in the $\lambda$ plane, the $n-1$ vacua are exchanged cyclically, and so we have vacuum crossing. Vacuum crossing here leads to the same consequences described in the Introduction; in particular, for any $n$ the theory with $\lambda=0$ flows to a CFT which is just the $\mathcal{N}=(2,2)$ $A_{n-1}$ minimal model (see \textit{e.g.} \cite{Witten:1993jg} and references therein). A similar analysis works in higher dimensions for any $n$, although when the superpotential is non-renormalizable we must think of these only as IR effective theories. 

As discussed in the Introduction, we can think of vacuum crossing as a limiting case of level crossing where the energy difference between the levels goes to zero. We now show this explicitly in these WZ theories.\footnote{The author thanks O. Mamroud for useful discussions on this example.} Consider adding a potential term which breaks both SUSY and the $\mb{Z}_{n-1}^R$ symmetry. For example, we can add a term $h|\phi-1|^2$ to the bosonic potential with very small $h$ (where $\phi$ is the bottom component of $\Phi$). This term splits the degeneracy between the $n-1$ vacua, and now we find a vacuum state and a low-lying state at energy $\Delta E\sim h$ above the vacuum. as we go around the circle in this deformed theory, the two states are exchanged, and so we have level crossing. Since the theory \eqref{eq:simplest_superpotential_2d} is obtained in the limit $h\to 0$, we find that vacuum crossing is indeed a limiting case of level crossing where the energy difference between the levels goes to zero, so that the "crossing" is confined to a specific energy level.

Next we discuss the relation to AISCs for these theories in even dimensions.\footnote{It is not clear yet if there is a relation for odd dimensions. One might expect a similar argument to work in odd dimensions using the parity anomaly, but the generalization is not obvious since the superpotential mass is not parity-odd.} We will do this explicitly for the $4d$ theory, but it is easy to generalize this to any even dimension. First we must find the AISC in these examples. We start by discussing the free $4d$ Weyl fermion $\psi$, which has a subtle AISC \cite{Cordova:2019jnf} which is not precisely of the form \eqref{eq:AISC}. Specifically, using spurion analysis we find that if we give $\psi$ a complex mass $M=e^{i\theta}m$ (where $m\in\mb{R}$), the partition function obeys
\begin{equation}
\frac{Z_\psi[e^{i\theta}m]}{Z_\psi[m]}=\exp\left(-\frac{i\theta}{384\pi^2}\int R\wedge R\right)\;,
\end{equation} 
where in our notation $\frac{1}{384\pi^2}\int R\wedge R\in \mb{Z}$ on closed spin manifolds. While there is no global AISC in the original sense here (since at $\theta=2\pi$ the phase is trivial on closed manifolds), there is a more subtle anomaly which prohibits us from identifying all of the points at $|m|\to \infty$. Let us now repeat this calculation for our $4d$ $\mc{N}=1$ WZ theories with superpotential \eqref{eq:simplest_superpotential_2d}. A similar spurion analysis shows that we actually do have an AISC in the standard sense of equation \eqref{eq:AISC} in these theories, since the partition function obeys
\begin{equation}\label{eq:WZ_AISC}
\frac{Z[e^{i\theta}\lambda]}{Z[\lambda]}=\exp\left(-\frac{i\theta}{384\pi^2}\frac{1}{n-1}\int R\wedge R\right) \;,
\end{equation}
and so the phase is not trivial for $\theta=2\pi$. The Witten index for these theories is $I=n-1$, and so due to this AISC we again find that there had to be vacuum crossing for this theory to be consistent.

\subsection{$4d$ $\mathcal{N}=1$ $SU(N)$ SQCD }\label{sec:4d_N1_YM}

Finally, we discuss vacuum crossing in a $4d$ $\mc{N}=1$ supersymmetric gauge theory. We focus on the case of an $SU(N)$ gauge group and $N_f<N$ flavors $Q_i,\tilde Q_i$ with identical mass $m$. Naively this theory has a $U(1)_R$ R-symmetry, but this symmetry is explicitly broken due to an ABJ anomaly to a $\mathbb{Z}_{2N}$ R-symmetry. Apart from the tree-level mass term, the effective potential is also corrected non-perturbatively  from the ADS superpotential (see e.g. \cite{Intriligator:1995au} for a review), so that the effective superpotential takes the form 
\begin{equation}
	W=m \tr M+(N-N_f)\left(\frac{\Lambda^{3N-N_f}}{\det M}\right)^{\frac{1}{N-N_f}}\;,
\end{equation}
where $M_{ij}=\tilde Q_i Q_j$ and $\lambda$ the strong coupling scale. This theory has $N$ gapped vacua, located at
\begin{equation}\label{eq:SQCD_vacua}
	M=c\id,\qquad\qquad c^N = \frac{\Lambda^{3N-N_f}}{m^{N-N_f}}\;.
\end{equation}
Thus the $\mathbb{Z}_{2N}$ R-symmetry is spontaneously broken to $\mathbb{Z}_2$. 

Next we write
\begin{equation}
\Lambda^{3N-N_f}=\mu^{3N-N_f}e^{2\pi i\tau}\;,
\end{equation} 
with $\tau$ the complexified gauge coupling, $\tau = \frac{4\pi i}{g^2}+\frac{\theta}{2\pi}$. We thus find that the $N$ vacua \eqref{eq:SQCD_vacua} are spaced out evenly on a circle in the complex plane, and their angle depends on $\theta$. We can ask what happens to the vacua as we take $\theta$ around in a loop to $\theta+2\pi $. It is easy to see that under this transformation the $N$ vacua are rotated cyclically into each other. We have thus found vacuum crossing. 

Due to the anomaly, we can consider the theory as a function of the complex mass term $|m|e^{i\theta}$. Then as we go around a circle in this parameter space, we find vacuum crossing. This means that the point $m=0$ must be singular. However, instead of a CFT, we find runaway vacua at this point. In other words, instead of the vacua merging, they disappear to infinity. We will discuss a similar example in Section \ref{sec:duality_free_field}.

\section{Applications Beyond Anomalies}\label{sec:applications}

We now discuss vacuum crossing more generally. In particular, we will encounter examples where vacuum crossing is not related to an anomaly by the simple mechanism described in the Introduction, and so we will study vacuum crossing independently of AISCs. We start by discussing two applications of vacuum crossing:
\begin{enumerate}
	\item Checking dualities.
	\item Looking for emergent symmetries.
\end{enumerate} 
Along the way, we will give more examples of vacuum crossing which are related to these applications. We then discuss a generalization of vacuum crossing, where we go through a phase transition as we go around the cycle in parameter space. We will use this generalization as another example of both of these applications.

\subsection{Checking Dualities}

If vacuum crossing occurs around some CFT for a specific theory, we also expect it to occur in any other theory which is IR dual to it. We can thus use the existence of vacuum crossing to check certain dualities. 

However, one must be careful when using this rule. In most of our examples, we cannot show that vacuum crossing occurs around specific CFTs, since these points are usually strongly coupled. Instead, the best we can do is show that it occurs far away from the CFT. It could be that there are many CFTs close to the origin of our parameter space, and it is enough for only one of them to exhibit vacuum crossing in order for the phase diagram to be consistent. Thus, even if vacuum crossing occurs far from the origin, the theory might flow to a CFT which does not exhibit vacuum crossing. In fact, it is possible that the vacua run off to infinity as we approach a specific CFT, so that in the IR we cannot even see the vacua that supposedly undergo vacuum crossing.

We now describe in detail a few examples where vacuum crossing occurs and can be matched across dualities. We then describe an example where it is not matched across a duality since vacuum crossing does not occur close enough to the CFT.

\subsubsection{A Four-Way Duality}\label{sec:4_way_duality}

We start by matching vacuum crossing across four theories which are conjectured to be dual. Consider the theory described in \cite{Gaiotto:2018yjh,Benini:2018bhk} and further analyzed in \cite{Rocek:2019eve}. This is a $3d$ $\mathcal{N}=1$ WZ model with a superfield $\Phi$ which is a traceless Hermitian $3\times 3$ matrix. The superpotential is
\begin{equation}\label{eq:3dWZmodelpotential}
W=\tr \Phi^3\;,
\end{equation}
and the symmetries are an $SU(3)$ symmetry with $\Phi$ in the adjoint representation, and a $\mathbb{Z}_2$ R-symmetry acting as $\Phi\to -\Phi$. This R-symmetry prohibits terms like $\tr \Phi^2$ from appearing. 

\begin{figure}
	\centering
	\includegraphics[scale=0.3]{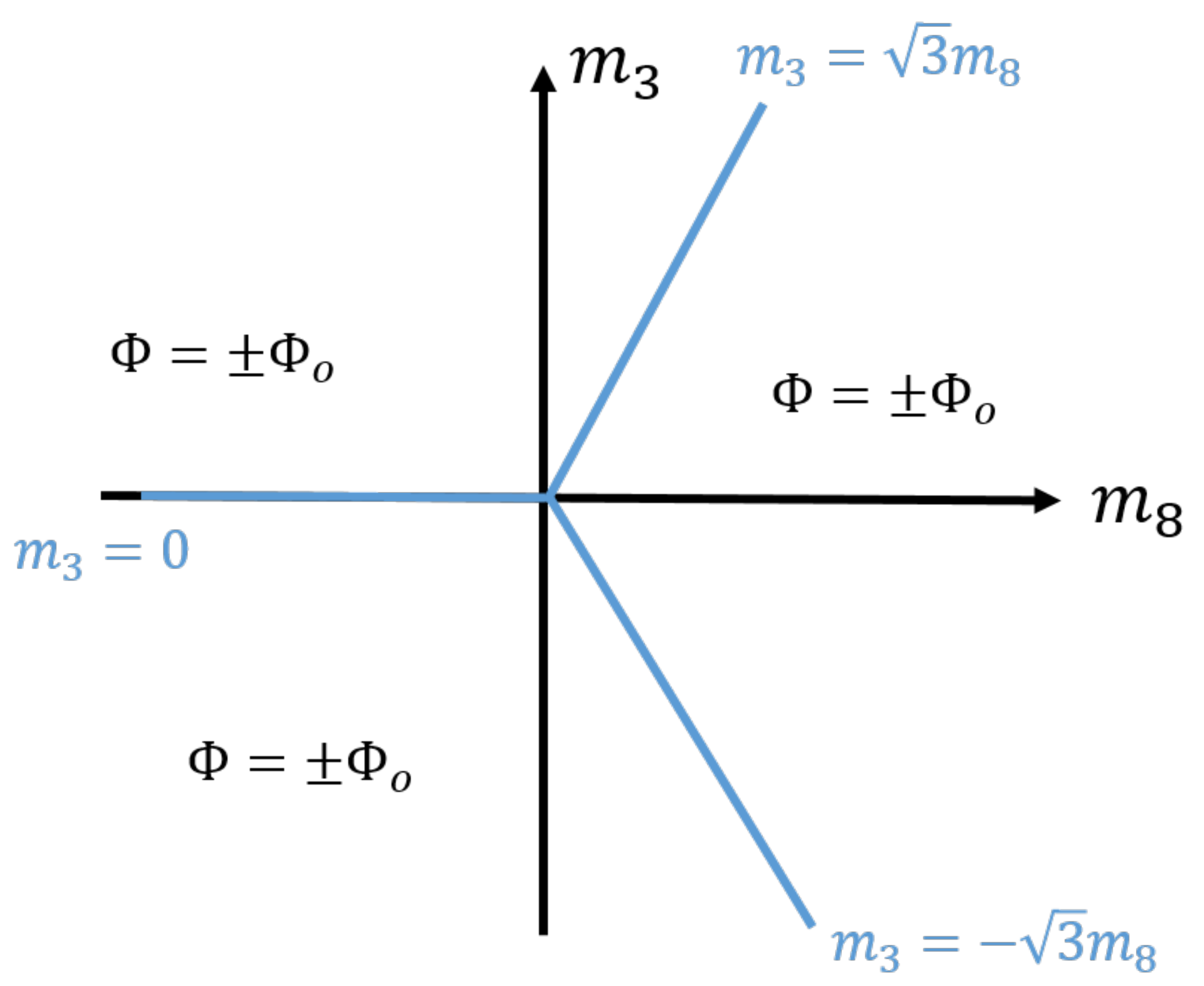}
	\caption{The phase diagram of the $3d$ $\mc{N}=1$ WZ model. At generic points we find two gapped vacua at low energies, while along the blue rays we find a $\mb{CP}^{1}$ model. $\Phi_0$ is given in Appendix \ref{sub:SU(3) WZ}, and a simple analysis shows that vacuum crossing occurs as we go around the origin.}
	\label{fig:3d_WZ_model}
\end{figure}

We now deform the theory by the additional term $\tr M \Phi$ with $M$ an Hermitian traceless matrix, explicitly breaking the $SU(3)$ global symmetry to its Cartan $U(1)\times U(1)$ (for generic values of $M$). We can use the global symmetry to diagonalize $M$ and bring it to the form $M=m_3 T_3+m_8 T_8$ with $T_a$ the generators of $SU(3)$ (normalized such that $\tr T_a T_b=\frac{1}{2}\delta_{ab}$). At large $m_3,m_8$ the theory is weakly coupled and one can solve for the vacua explicitly. One finds that for generic values of $m$, there are two gapped vacua at $\Phi=\pm \Phi_0$, so that the $\mb{Z}_2$ R-symmetry is spontaneously broken. In addition, there are three rays that start at the origin along which the effective theory is a $\mb{CP}^1$ model.
In Appendix \ref{sub:SU(3) WZ} we find the full phase diagram of this theory, and show that it exhibits vacuum crossing between the two gapped vacua. We present this phase diagram in Figure \ref{fig:3d_WZ_model}.

This example is different from the ones discussed above, since a moduli space opens up as we go around the loop in parameter space. This means that we can no longer argue that vacuum crossing must occur if the theory has an anomaly and level crossing is prohibited, and so vacuum crossing here might not be related to an anomaly. In addition, the appearance of a moduli space means that we can continuously go around the cycle and end up at the same vacuum, and so we cannot argue that there exists a CFT at the origin (however, in these examples we can use the phase structure itself to argue for the existence of a CFT near the origin, since the point where the three moduli spaces meet is a singularity). However, there is still some form of vacuum crossing occurring here. This is a result of a $\mathbb{Z}_3$ symmetry which rotates the phase diagram \ref{fig:3d_WZ_model} by $120$ degrees, and allows us to identify the vacua on each side of a $\mathbb{CP}^1$ line by performing a symmetry transformation taking us from the region before the $\mathbb{CP}^1$ line to the region after it. We can thus follow a "specific" vacuum as we go around a cycle in parameter space, and check where it ends up. This form of vacuum crossing is still useful since dual theories must match it.

This WZ model has a few conjectured IR duals\footnote{The recent bootstrap analysis \cite{Rong:2019qer} found some possible issues with the duality between the WZ theory and theory (a). These must be studied more carefully in order to make the statement of the duality more precise.} \cite{Gaiotto:2018yjh,Fazzi:2018rkr,Choi:2018ohn}:
\begin{enumerate}[(a)]
	\item 
	An $\mathcal{N}=2$ $U(1)_0$ gauge theory with two matter fields of charge $1$.
	\item An $\mathcal{N}=2$
	$SU(3)_{5/2}$ gauge theory with three matter fields $Q_i$ and superpotential $W=\epsilon^{abc}\epsilon_{ijk}Q_{a}^{i}Q_{b}^{j}Q_{c}^{k}$.
	\item An $\mathcal{N}=2$ $U(2)_{2,-2}$ gauge theory with 2 fundamental matter fields.
\end{enumerate}
As a check of these dualities, one can find whether they all exhibit vacuum crossing when we add the appropriate couplings. Indeed, we show this for theories (a),(b) in Appendix \ref{app:4_way_duality}, while for theory (c) this can be shown using similar arguments \cite{SahandUnpublished}. So we indeed find that vacuum crossing is matched across all of these theories, which is consistent with the conjectured dualities.

Note that two of the dual theories above do not have a UV time-reversal symmetry which relates the two vacua. However, the duality tells us that such a symmetry must appear in the IR. We will discuss the relation between vacuum crossing and these emergent symmetries in Section \ref{sec:emergent_syms}.

As an aside, we discuss the $2d$ version of the WZ model \eqref{eq:3dWZmodelpotential}.\footnote{The author thanks Z. Komargodski for useful discussion on the $2d$ version of the WZ model.} This is a $2d$ $\mathcal{N}=(1,1)$ WZ model with a single chiral superfield in the adjoint of $SU(3)$ and the same superpotential described above:
\beq \label{eq:2d_superpotential} W=\tr \Phi^3+\tr M\Phi\;. \eeq 
While the $3d$ version of this model was studied in the literature, the $2d$ version has not been discussed. At $M=0$ the theory has a $\mathbb{Z}_2$ R-symmetry with similar consequences to those discussed in \cite{Gaiotto:2018yjh}, and for general $M$ its classical phase diagram is identical to the $3d$ phase diagram in Figure \ref{fig:3d_WZ_model}. However, since there are no moduli spaces in $2d$, the $\mathbb{CP}^1$ model appearing along the three rays is lifted by quantum effects. Since the $\mathbb{CP}^1$ model has Witten index 2, we expect there to remain two trivial vacua, and so the low energy theories along these rays are identical to the rest of the phase diagram. The entire phase diagram thus consists of two gapped vacua. Just like in $3d$, vacuum crossing occurs as we go around the origin, but since no moduli spaces appear in $2d$ this leads to stronger constraints on the theory. In particular, since vacuum crossing occurs, we find that there must be a CFT near the origin. Presumably this CFT is at $M=0$, since the theory has enhanced symmetry there.

We can ask what CFT appears at the origin of the $2d$ model. Just like the $3d$ model, there are signs that this CFT has emergent $\mc{N}=2$ SUSY. Specifically, one can repeat the analysis of \cite{Rocek:2019eve} for the $2d$ theory. The calculation is almost identical to the $3d$ case, and the result is that the domain walls between the two gapped vacua have emergent SUSY with 2 supercharges. Assuming these are 1/2-BPS, this is consistent with emergent SUSY with $4$ supercharges appearing at the CFT. The vacuum structure is also consistent with emergent $\mc{N}=2$ SUSY. However, these checks are relatively weak and so further evidence is required for the emergent SUSY. In particular, it would be nice to find a dual description, possibly with $\mc{N}=2$ SUSY. Unfortunately we have not found such a theory yet. Presumably, this dual should also exhibit vacuum crossing.

\subsubsection{$2d$ Mirror Symmetry}

We consider the mirror duals of the two theories discussed in Section \ref{sec:2d_U1_gauge_theory}. Since we showed that both theories exhibit vacuum crossing, we expect it to occur in their duals as well.

The first theory was a $2d$ $\mc{N}=(2,2)$ $U(1)$ gauge theory with $N_f$ superfields of charge 1. Its mirror dual has twisted superpotential (see \cite{Hori:2000kt} and references therein)
\begin{equation} \widetilde{W}=\frac{1}{4\pi}\left[\Sigma\left(\sum_{k=1}^{N_f} Y_k+2\pi i t\right)+i c \sum_{k=1}^{N_f} e^{-Y_k}\right]\;,  
\end{equation}
where $t$ is the complexified FI parameter and $c$ is a constant. Here $\Sigma$ is the twisted chiral multiplet corresponding to the field strength of the vector multiplet of the original theory, and $Y_k \sim Y_k+2\pi i $ are also twisted chiral multiplets. Solving for the vacua of the theory, we find $N_f$ vacua where the $Y_k$'s all have the same vev
\begin{equation} Y_1=...=Y_{N_f}\equiv Y_{(j)}\;,
\end{equation}
with 
\begin{equation} 
Y_{(j)}=-\frac{2\pi i(t+j)}{N_f},\;\;\;\;\;j=1,...,N_f\;.
\end{equation}
It is then easy to see that taking $t$ in a loop to $ t+1$ leads to vacuum crossing as expected. 

Next we consider the $U(1)$ gauge theory with one field of charge $q$. Its mirror dual has twisted superpotential \cite{Hori:2000kt}
\begin{equation} \widetilde{W}=\frac{1}{4\pi}\left[\Sigma\left(q Y+2\pi i t\right)+ic e^{-Y}\right]\;,  
\end{equation}
with $c$ some constant. There are thus $q$ vacua at 
\begin{equation}  Y_{(j)}=-\frac{2\pi i(t+j)}{q}\;,\;\;\;\;\;j=1,...,q. \end{equation}
Once again we find that vacuum crossing occurs, as expected.

\subsubsection{Duality Between a $4d$ $\mathcal{N}=1$ WZ theory and a Free Field}\label{sec:duality_free_field}

We now discuss an example where vacuum crossing is not matched across a duality, and it is interesting to understand why this fails. The duality is between a $4d$ $\mc{N}=1$ WZ theory with superpotential $W=\frac13\Phi^3$ and a free chiral multiplet.\footnote{Note that the WZ model is IR free, so we treat it as an IR effective theory and only consider it for small values of $\Phi$ where it is valid.} As discussed in Section \ref{sec:WZ_model_w_4_supercharges}, the WZ theory exhibits vacuum crossing when we deform it by the superpotential term $\lambda\Phi$. However, the dual theory does not have a deformation which exhibits this phenomenon. In fact, there is no dual deformation which creates the same vacuum structure for this theory. The reason is that as we go to low energies in the WZ model, the two vacua go to infinity in $\Phi$ space (since the coefficient of $\Phi^3$ goes to zero). These vacua thus do not appear in the corresponding deformation in the dual theory, and in particular, vacuum crossing does not occur.

We have thus found an example where vacuum crossing occurs far from the CFT on one side of a duality and is not matched on the other side. Another mechanism where this can happen is if there are many CFTs close to the origin, in which case vacuum crossing may occur only around some of these CFTs and not around others. We conclude that in general one must be careful when matching vacuum crossing across dualities if the vacuum crossing occurs far away from the CFT.

\subsection{Emergent Symmetries}\label{sec:emergent_syms}

In most of the examples of vacuum crossing discussed above, the vacua which are exchanged are related by some broken symmetry. This is satisfying due to the mechanism discussed in the Introduction in which vacuum crossing occurs due to a broken symmetry and an AISC. 
However, some of the examples seem to exhibit vacuum crossing without a symmetry which relates the vacua. Interestingly, it turns out that in all such examples above, the situation is a bit more subtle; while there is no UV symmetry which relates the vacua, there is an emergent IR symmetry which appears near the CFT which does relate them. We thus find that there \textit{is} a symmetry relating the vacua, but it is only visible in the IR, near the CFT. We now show this explicitly.

Consider the four-way duality from Section \ref{sec:4_way_duality}.  As emphasized in \cite{Fazzi:2018rkr}, the full symmetry of the IR theory is not manifest in any of the UV descriptions; the CFT has $\mc{N}=2$ SUSY, an $SU(3)$ global symmetry and time-reversal symmetry, but each of the UV descriptions exhibits at most only two of these three. Let us focus on the theories (b),(c) of Section \ref{sec:4_way_duality}, which do not have time-reversal symmetry in the UV. The phase diagram of these theories consists of two gapped vacua which undergo vacuum crossing, and there is no UV symmetry which relates them. However, from the duality we know that these vacua are related by an emergent time-reversal symmetry in the IR. So the existence of vacuum crossing is related to the appearance of an emergent IR symmetry which relates the two vacua.

A similar result appears in the generalized version of vacuum crossing to be discussed in Section \ref{sec:a_generalization}, where there exists a phase transition as we go around the cycle in parameter space. We will find more theories which exhibit vacuum crossing without a symmetry which relates the vacua, but again, a duality will tell us that there exists an emergent discrete symmetry in the IR which relates the vacua.

We have thus seen that in every example discussed above, vacuum crossing is related to the existence of a spontaneously broken symmetry (which might be emergent). One can ask whether this is a generic phenomenon.\footnote{The author thanks O. Aharony for useful discussions on this subject.} Indeed, it seems as though it is not possible to find a simple SUSY WZ model with 4 supercharges in which vacuum crossing is not related to a symmetry. To see this, consider a $2d$ $\mc{N}=(2,2)$ WZ model with a single chiral superfield $\Phi$ (the example can be immediately generalized to higher dimensions). If the superpotential is cubic then by a redefinition of the superfield we can always bring it to the form discussed in Section \ref{sec:WZ_model_w_4_supercharges}, and so vacuum crossing is related to a broken $\mb{Z}_2$ symmetry. We thus consider a general quartic superpotential (after shifting $\Phi$ to remove the cubic term):
\beq
W=\frac{1}{4}\Phi^4+\frac{\lambda_2}{2}\Phi^2+\lambda_1 \Phi\;.
\eeq
For vacuum crossing, we need the parameter space to be 1 complex-dimensional, so we consider some codimension 1 surface in the $(\lambda_1,\lambda_2)$ plane. For concreteness we will fix $\lambda_2=1$, but due to the topological nature of vacuum crossing we do not expect it to depend on small changes in the couplings. The F-term equations are cubic equations, and so we generically find 3 SUSY vacua with singularities wherever two of the vacua meet, and around these singularities we find vacuum crossing. The full theory does not have any global symmetry which can relate the two vacua, but expanding around the singularities we find an emergent symmetry which does relate them. Indeed, expanding the superpotential around a singularity we find that it should be described by the effective superpotential $W\propto \Phi^3$, which reduces to the same theory described in Section \ref{sec:WZ_model_w_4_supercharges}. In particular, there appears an emergent $\mb{Z}_2$ symmetry relating the two vacua which meet at the singularity, and so vacuum crossing here is related to an emergent symmetry. This analysis is easily generalized to superpotentials with higher powers of $\Phi$, and to cases where $n>2$ vacua meet at a singularity (in which case the effective theory will be $W\propto\Phi^{n+1}$, and we find an emergent $\mb{Z}_{n}$ R-symmetry). Thus, vacuum crossing is always related to a (possibly emergent) broken symmetry in these SUSY WZ theories.

Indeed, it is intuitively clear why vacuum crossing must be related to a symmetry as long as we do not cross a phase transition or a moduli space as we go around the loop in parameter space. As we approach the singular point where the vacua meet, the loops in parameter space along which we find vacuum crossing get smaller and smaller. We thus find that by a smaller and smaller change of our parameters, we can map one vacuum into the other. We thus expect there to appear an approximate symmetry relating the vacua as we approach the singularity. Since theories with holomorphic superpotentials do not have phase transitions (since phase transition lines must be complex codimension 1, so that we can always "go around" them), this explains why we cannot find a simple complex WZ theory which is a counterexample.

These arguments and examples lead us to the following conjecture. If vacuum crossing occurs close enough to a CFT, and if we do not cross a phase transition or a moduli space as we go around the loop in parameter space, then this vacuum crossing must be related to the existence of a spontaneously broken symmetry. In particular, if the symmetry is not visible in the UV, it must be emergent in the IR. This gives us a tool to look for IR emergent symmetries in strongly coupled theories. 

There are signs that an even stronger claim can be made. Indeed, we have discussed the example of the four-way duality, where vacuum crossing is related to an emergent symmetry even though we cross a moduli space. In addition, in the next section we will see examples where this occurs even though we go through a phase transition. One can then ask whether vacuum crossing must be related to a broken symmetry even if we cross a moduli space or a phase transition. However, this claim seems too strong.\footnote{One can imagine a case where vacuum crossing occurs between two vacua which have completely different low-energy theories, but which "transform into each other" as we cross a phase transition. Since the vacua are different, there cannot be a symmetry relating them.} Thus, at the moment it is unclear whether a definitive statement can be made regarding the relation between vacuum crossing and symmetries in cases where the loop in parameter space crosses a phase transition or a moduli space.

\subsection{A Generalization}\label{sec:a_generalization}

Let us discuss a generalization of vacuum crossing, in which there is a phase transition as we vary the periodic parameter $\theta$. This example will also give us another example of matching vacuum crossing across dualities, and of the relation between vacuum crossing and emergent symmetries.

\begin{figure}
	\centering	
	\includegraphics[scale=0.35]{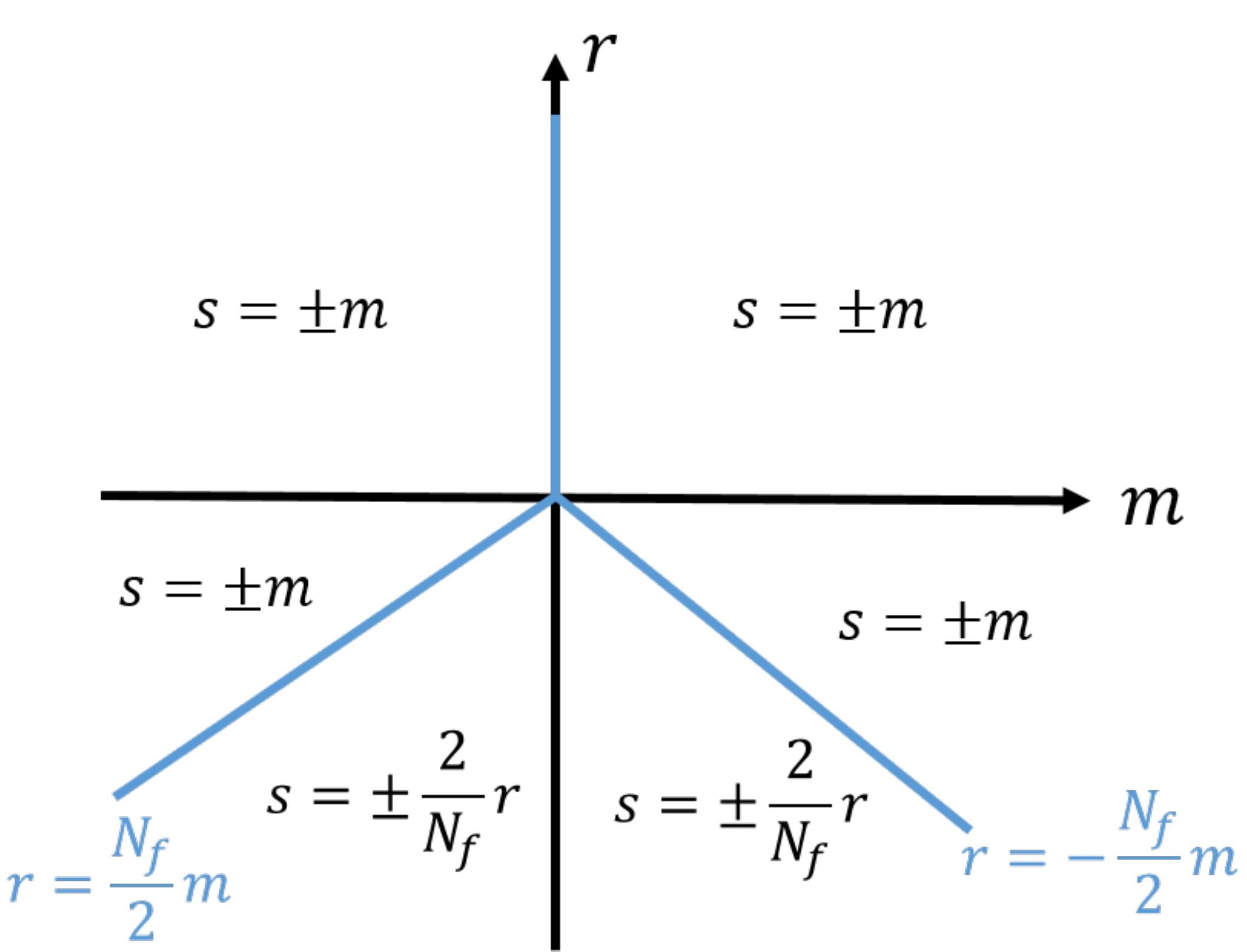}
	\caption{Phase diagram of the $U(1)_0$ gauge theory with $N_f$ charge $+1$ matter fields (we draw the phase diagram at generic points and ignore some rays along which moduli spaces open up). The low-energy theories at $s=\pm\frac{2}{N_f}r$ are $U(1)_{\mp N_f/2}$, while the low-energy theories at $s=\pm m$ are a $\mb{CP}^{N_f/2-1}$ model. Along the blue lines, moduli spaces open up. As we go around the circle, we find vacuum crossing.}
	\label{fig:Generalization}
\end{figure}

We discuss a $3d$ $\mc{N}=2$ $U(1)_0$ gauge theory with $N_f$ fields $Q_i$ of charge $+1$ (the case $N_f=2$ was discussed in Section \ref{sec:4_way_duality}).\footnote{Note that our conventions for the Chern-Simons level force $N_f$ to be even.} The theory has a global $SU(N_f)\times U(1)_T$ global symmetry. We deform the theory by an FI term $r$, and also by real masses $M_{ij}$ for the $SU(N_f)$ symmetry which take the form
\begin{equation}
M_{ij}=m\begin{pmatrix}
\id_{\frac{N_f}{2}\times\frac{N_f}{2}} &\\
& -\id_{\frac{N_f}{2}\times\frac{N_f}{2}}
\end{pmatrix}\;,
\end{equation}
for some $m\in\mb{R}$ and where $\id_{k\times k}$ is the $k\times k$ identity matrix. This choice preserves an $S\left[U\left(N_f/2\right)\times U\left(N_f/2\right)\right]$ global symmetry, and so while $m$ can change due to quantum corrections, the structure of $M_{ij}$ will not. We solve for the vacuum structure of this theory in Appendix \ref{app:3d_N=2_U(1)_Nf}, and present the result in Figure \ref{fig:Generalization}. Denoting the bottom component of the vector multiplet by $s$, 
we find that the low-energy theories for the vacua at $s=\pm\frac{2}{N_f}r$ are $U(1)_{\mp N_f/2}$, while for the vacua at $s=\pm m$ we find $\mb{CP}^{N_f/2-1}$ theories. We have only described the vacua at generic points; there are some special lines where moduli spaces open up.

We can now study vacuum crossing in this example. Following one of the vacua around a circle which goes around the origin in Figure \ref{fig:Generalization}, we find that we end up at the other vacuum. However, these examples are different than those discussed above, since the effective theories at the vacua change as we go around the circle due to phase transitions. As a result of these phase transitions (and since moduli spaces appear as we go around the circle), we cannot argue that vacuum crossing is related to an anomaly here. In addition, we cannot use vacuum crossing to argue for the existence of a CFT at the origin here (although once again this can be done using other methods). However, we can still discuss the matching of this version of vacuum crossing across dualities. Conjectured dual theories include the following $3d$ $\mc{N}=2$ gauge theories \cite{Choi:2018ohn,Aharony:1997bx,Giveon:2008zn,Benini:2011mf,Aharony:2014uya}:
\begin{itemize}
	\item $U(N_f/2-1)_0$ with $N_f$ fundamental superfields,
	\item $U(N_f/2+1)_{-2,N_f}$ with $N_f$ fundamental superfields,
	\item $SU(N_f/2)_1$ with $N_f$ fundamental superfields.
\end{itemize}
Indeed, one can check that the simple cases of these duals also exhibit vacuum crossing. 

Finally, we discuss emergent symmetries in these examples. Note that for the last two dual theories, time-reversal is not a symmetry in the UV, and so the vacua are not related by any symmetry. However, the duality tells us that in the IR an emergent time reversal symmetry appears. We thus have many more examples where vacuum crossing is related to the appearance of an emergent IR symmetry, even though we cross a phase transition as we go around the loop in parameter space.

\section{Summary}

In this note we discussed vacuum crossing, which occurs when vacua are exchanged nontrivially as we continuously take a periodic parameter $\theta$ in a loop from $\theta$ to $\theta+2\pi$. We showed that in some cases this phenomenon is related to AISCs, and also discussed applications beyond anomalies. In particular, we argued that vacuum crossing can be used to check dualities, and that it can be used to search for emergent symmetries. This makes vacuum crossing a useful new tool to study strongly coupled QFTs.

There are some generalizations which we did not consider in this note. Specifically, in all of our examples we considered parameter spaces which were two-dimensional. We can ask what consequences of vacuum crossing can be generalized to cases where the parameter space $\mc{M}$ is $d$-dimensional for $d>2$. Assuming that all of our vacua come from some broken discrete symmetry group $G$, we can think of our space of vacua as being $G$-valued at every point on $\mc{M}$. We thus have a principal $G$-bundle with base space $\mc{M}$. It turns out that since $G$ is discrete, these bundles are trivial unless $\pi_1(\mc{M})\neq 0$. In other words, unless $\pi_1(\mc{M})\neq 0$, we can always "unwind" the vacuum crossing structure, meaning that there does not have to be a singularity at the origin. Indeed, in all of our examples above we think of our parameter space as being $\mc{M}=\mathbb{R}^2/\{x_0\}$ where $x_0$ is the position of our singularity, and so $\pi_1(\mc{M})$ is not trivial. In higher dimensions it is just technically more complicated to find cases where $\pi_1(\mc{M})\neq 0$, and where our analysis holds. One example where this occurs is if there exists a conformal manifold in our $d$-dimensional parameter space. For example, consider the $3d$ $\mathcal{N}=2$ WZ theory with three superfields $X,Y,Z$ and superpotential
\begin{equation}
W=\tau XYZ -\left(X^3+Y^3+Z^3\right)-\lambda X\;.
\end{equation}
For $\lambda=0$, $\tau$ is an exactly marginal operator \cite{Strassler:1998iz} (for a recent analysis see \cite{Baggio:2017mas}). The parameter space is thus\footnote{The parameter space is more subtle since there is a duality group which acts on $\tau$, but this does not affect our main points.} $\mc{M}=\mathbb{C}^2/\mathbb{C}$, since we have a singularity at $\lambda=0$ for all $\tau$. In particular, we find $\pi_1(\mc{M})\neq 0$. A quick analysis shows that vacuum crossing does occur as we take $\lambda\to e^{2\pi i}\lambda$, and since $\pi_1(\mc{M})\neq 0$ we find that this leads to nontrivial consequences.\footnote{Of course, this example is weaker than the previous ones, since we had to know in advance that there is a CFT for each $\tau$ in order to get nontrivial results.}

One can also consider vacuum crossing when the theory has a continuous moduli space of vacua. First we discuss the case where this moduli space is a result of the breaking of a continuous symmetry. It is not clear that one can clearly define vacuum crossing in such a case, since all of the points are equivalent and so one can always continuously rotate the moduli space to remove any "vacuum crossing" in it. Thus it does not seem as though vacuum crossing is useful in theories with a broken continuous symmetry. However, if the moduli space is not a result of a broken symmetry, then it could have "special points" which are distinguishable from other points (\textit{e.g.} by the effective theory at that point). In this case we expect the consequences to be similar to vacuum crossing if these special points have nontrivial winding numbers as we go around cycles in parameter space. 

We have only discussed a few examples of vacuum crossing and its applications in this note. We hope that vacuum crossing will find more uses in the future.

\section*{Acknowledgments}

The author is grateful to C. Choi, A. Giveon, R. Kalloor, K. Roumpedakis,  T. Sheaffer, E. Urbach, Y. Wang and M. Watanabe for helpful discussions. The author would especially like to thank O. Aharony, Z. Komargodski, H. T. Lam, O. Mamroud and L. Tizzano for discussions and for comments on a draft of the paper. The author also thanks S. Seifnashri for discussions and for sharing unpublished work. Finally, the author is grateful to the SCGP for hospitality during the initial stages of this project. The author is supported by an Israel Science Foundation center for excellence grant and by the I-CORE program of the Planning and Budgeting Committee and the Israel
Science Foundation (grant number 1937/12).

\newpage

\begin{appendices}
\addtocontents{toc}{\protect\setcounter{tocdepth}{1}}

\section{Vacuum Crossing in the Four-Way Duality}\label{app:4_way_duality}

\subsection{$\mc{N}=1$ WZ Model with $SU(3)$ Symmetry\label{sub:SU(3) WZ}}

We consider the $3d$ $\mathcal{N}=1$ WZ model with a single superfield $\Phi$ in the adjoint of $SU(3)$ and superpotential
\begin{equation}
W=\tr \Phi^3+\tr M\Phi\;.
\end{equation}
Using the global $SU(3)$ symmetry to diagonalize $M$ and bring it to the form $M=m_3 T_3+m_8 T_8$ with $T_a$ the generators of $SU(3)$ (normalized such that $\tr T_a T_b=\frac12\delta_{ab}$), we find that at generic values of $m_3,m_8$ there are two vacua at $\Phi=\pm \Phi_0$. Explicitly, writing $\Phi_0=\phi_0^a T_a$ we find:
\begin{equation}\label{eq:3dWZmodelvacua}
\begin{split}
\phi_0^3&=-\sqrt[4]{3} \sqrt{\sqrt{m_3^2+m_8^2}-m_8}\\
\phi_0^8&=\frac{\sqrt[4]{3} \sqrt{\sqrt{m_3^2+m_8^2}-m_8} \left(\sqrt{m_3^2+m_8^2}+m_8\right)}{m_3}\;,\\
\end{split}
\end{equation}
with all other $\phi_0^a$'s vanishing.
In addition, there are three rays that start at the origin along which the effective theory is a $\mb{CP}^1$ model. 

The full phase diagram appears in Figure \ref{fig:3d_WZ_model}. One can now slowly move in a circle in the $(m_1,m_2)$ plane and see how the vacua \eqref{eq:3dWZmodelvacua} change. Doing this, we find that the vacua are exchanged after we complete one rotation in the $(m_1,m_2)$ plane. We thus have vacuum crossing.

\subsection{$3d$ $\mathcal{N}=2$ $U(1)_{0}$ with Two Fields of Charge 1}\label{sub:3d QED with two quarks}

Consider a $3d$ $\mathcal{N}=2$ $U(1)_{0}$ with two fields of charge $+1$. Following \cite{Intriligator:2013lca}, we can find the vacua by solving the F-term equations in the low-energy effective superpotential for the vevs of the matter fields $Q_1,Q_2$ and of the adjoint scalar $s$. The vacua are found in Appendix \ref{app:3d_N=2_U(1)_Nf}, and we find that for generic values of the parameters we have two solutions $s_{\pm}$. The phase diagram appears in Figure \ref{fig:Generalization}, and we find that we have vacuum crossing as we go around the loop in parameter space. Note that unlike most of the examples in Appendix \ref{app:3d_N=2_U(1)_Nf}, in this case we do not cross a phase transition as we go around the loop.

\subsection{A $3d$ $\mathcal{N}=2$ $SU(3)$ Gauge Theory}\label{sub:SU(3)_gauge_theory}

\begin{figure}
	\centering
	\includegraphics[width=0.4\linewidth]{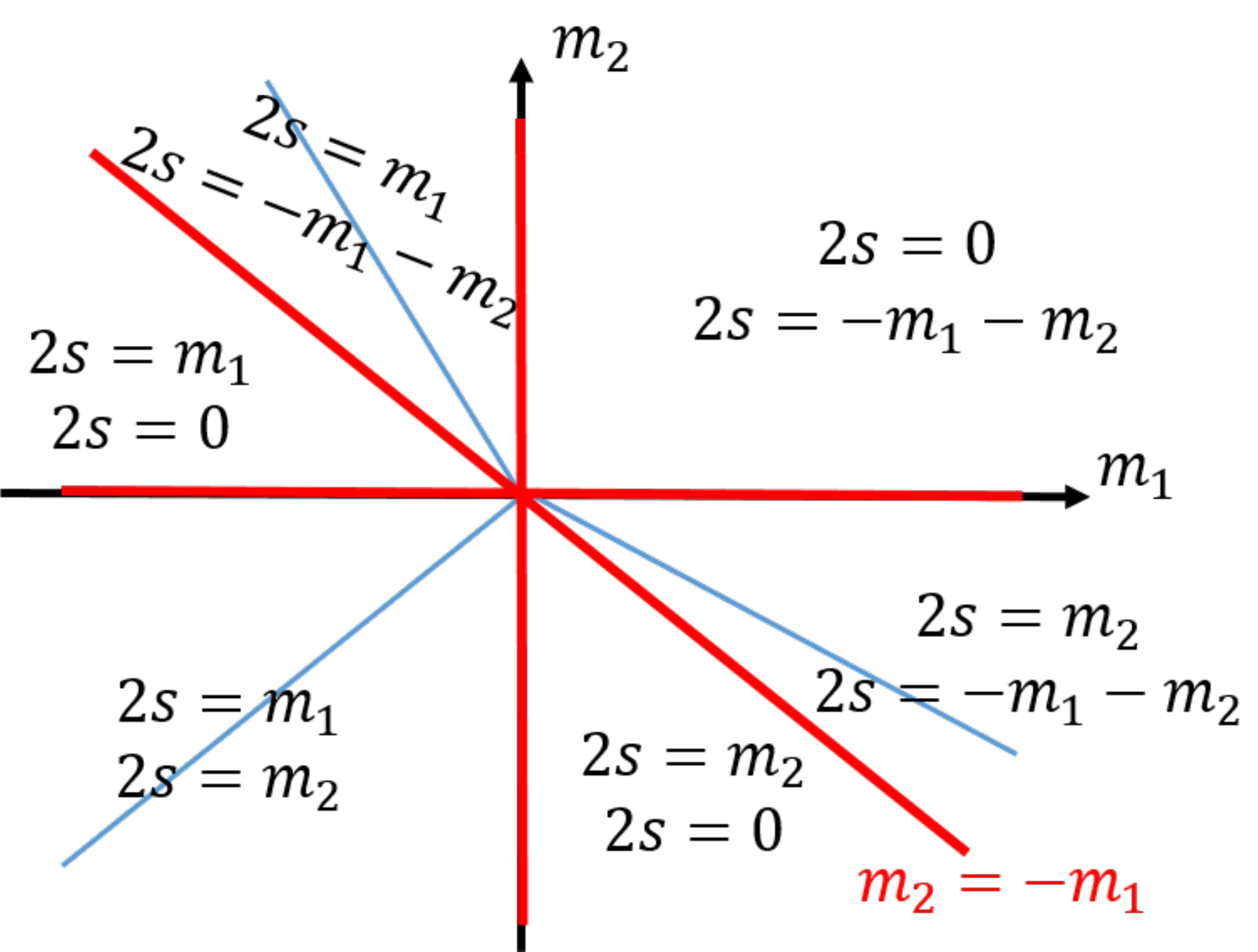}
	\caption{Phase Diagram for the $\mc{N}=2$ $SU(3)_{5/2}$ gauge theory. The vacua at generic points are trivial theories, while blue lines indicate moduli spaces. The solutions "jump" along the red lines. The transitions along the red lines aren't smooth, as opposed to the transitions in the dual theories. We expect these transitions to be smoothed out due to quantum corrections.}
	\label{fig:SU(3)_gauge}
\end{figure}

We now consider a $3d$ $\mc{N}=2$  $SU(3)_{5/2}$ gauge theory with $N_{f}=3$ fundamental fields $Q_i$ and a superpotential $W=\epsilon^{abc}\epsilon_{ijk}Q_{a}^{i}Q_{b}^{j}Q_{c}^{k}$. We deform the theory by a real mass matrix for the $SU(3)$ global symmetry, with real masses $m_1,m_2$ and $m_{3}=-m_1-m_2$. Once again we can solve for the vacua of the theory as a function of the masses. The vacua were found in \cite{Rocek:2019eve}, where it was found that vacua exist only when the adjoint scalar $s$ in the vector multiplet takes the form $s=\text{diag}(\sigma,\sigma,-2\sigma)$. Apart from some rays emanating from the origin, there are exactly two solutions for the vacuum, with $\sigma$ taking two of the possible values $\sigma\in\{0,m_1,m_2,m_3\}$. All of these SUSY vacua turn out to be trivial theories. Explicitly, the phase diagram appears in Figure \ref{fig:SU(3)_gauge}.
Following the vacua as we go around a large loop in parameter space, we find that they are indeed exchanged as we go around the origin. We thus have vacuum crossing.

\section{Vacua of the $3d$ $\mathcal{N}=2$ $U(1)_0$ Theory with $N_f$ Fields of Charge 1}\label{app:3d_N=2_U(1)_Nf}

We study a $3d$ $\mc{N}=2$ $U(1)_0$ gauge theory with $N_f$ fields $\vec{Q}$ of charge 1, where the real mass matrix is taken to be
\begin{equation}
M=m\begin{pmatrix}
\id_{\frac{N_f}{2}\times\frac{N_f}{2}} &\\
& -\id_{\frac{N_f}{2}\times\frac{N_f}{2}}
\end{pmatrix}\;,
\end{equation}
for some $m\in\mathbb{R}$. This choice of $M$ preserves an $S[U(N_f/2)\times U(N_f/2)]$ symmetry. Following \cite{Intriligator:2013lca}, the vacuum equations are
\begin{align*}
\left|Q_i\right|^2+\left|Q_j\right|^2 & =r+\frac{1}{2}\frac{N_f}{2}\left(	|s+m|+|s-m|\right)\\
(s+m)Q_i & =0\\
(s-m)Q_j & =0\;,
\end{align*}
where $s$ is the scalar component of the vector multiplet, and where $i=1,...,\frac{N_f}{2}$ and $j=\frac{N_f}{2}+1,...,N_f$. The solutions are:
\begin{itemize}
	\item If $s=-m\neq0$ then $Q_j=0$ and $|Q_i|^2=r+\frac{N_f}{2}|m|$, which can only happen for $r+\frac{N_f}{2}|m|>0$. The low-energy theory is a $\mathbb{CP}^{N_f/2-1}$ model.
	\item If $s=m\neq 0$ then $Q_i=0$ and $|Q_j|^2=r+\frac{N_f}{2}|m|$, which can only happen for $r+\frac{N_f}{2}|m|>0$. The low-energy theory is again a $\mathbb{CP}^{N_f/2-1}$ model.
	\item If $s=m=0$ then we find a $\mb{CP}^{N_f-1}$ moduli space of vacua for $r>0$, corresponding to the solutions of $|Q_i|^2+|Q_j|^2=r$.
	\item If $Q_i=Q_j=0$ then if $r<-\frac{N_f}{2}|m|$ we have vacua at $s=\pm \frac{2}{N_f}r$ which have a $U(1)_{\mp \frac{N_f}{2}}$ low-energy theory. In addition, along the special lines $r=\pm \frac{N_f}{2}m$ a Coulomb branch opens up for $r<0$. We will not worry about these lines.
\end{itemize}
The phase diagram is presented in Figure \ref{fig:Generalization} (we focus on vacua at generic points, and do not dicsuss the moduli spaces that open up along some rays). It is simple to see that as we go around in a circle in parameter space, the vacua are exchanged, and so we find vacuum crossing.
\end{appendices}

\bibliographystyle{./aug/ytphys}
\bibliography{refs}

\end{document}